\documentclass{article}
\usepackage{spconf,amsmath,graphicx}
\usepackage{url}
\usepackage{hyperref}
\usepackage{enumitem}

\title{GE2E-KWS: Generalized End-to-End Training and Evaluation for Zero-shot Keyword Spotting}
\name{Pai Zhu, Jacob W. Bartel, Dhruuv Agarwal, Kurt Partridge, Hyun Jin Park, Quan Wang}
\address{Google LLC,
Mountain View, CA, U.S.A\\
\texttt{\{paizhu,bartel,dhruuv,kep,hjpark,quanw\}@google.com}}

\begin{document}
%
\maketitle
\begin{abstract}

We propose GE2E-KWS---a generalized end-to-end training and evaluation framework for customized keyword spotting. Specifically, enrollment utterances are separated and grouped by keywords from the training batch and their embedding centroids are compared to all other test utterance embeddings to compute the loss. This simulates runtime enrollment and verification stages, and improves convergence stability and training speed by optimizing matrix operations compared to SOTA triplet loss approaches. To benchmark different models reliably, we propose an evaluation process that mimics the production environment and compute metrics that directly measure keyword matching accuracy. Trained with GE2E loss, our 419KB quantized conformer model beats a 7.5GB ASR encoder by 23.6\% relative AUC, and beats a same size triplet loss model by 60.7\% AUC. Our KWS models are natively streamable with low memory footprints, and designed to continuously run on-device with no retraining needed for new keywords (zero-shot).

\end{abstract}

\section{Introduction}
\label{sec:intro}

Traditionally, the goal of keyword spotting(KWS) is to detect a relatively small set of predefined keywords such as device-waking phrases like ``Hey Google'' or simple control commands like ``stop'', ``turn on lights'', etc in a streaming audio environment. Research usually focuses on developing small-footprint models that can run continuously on device~\cite{chen2014small,alvarez2019end,zhu2023locale}. As devices become more intelligent and personalized, there is a growing demand from customers for the flexibility to specify their own triggering keyword via text or audio. A custom keyword spotting solution can provide a natural and seamless way for users to activate devices or software using voice.


Custom keyword spotting is able to detect a custom target phrase from audio streams.  While researchers have tried to use template matching approaches such as Dynamic time warping(DTW)~\cite{barakat2011keyword} and phoneme classification approaches~\cite{lopez2021deep}, they pose limitations on the keyword choices or require retraining for new keywords. A more efficient way to achieving open-vocabulary and zero-shot KWS is by 1) representing utterances by speech embeddings, and 2) comparing a speech embedding generated at enrollment time with a speech embedding generated at serving time. Detection of the keyword is triggered only when the cosine similarity of the two speech embeddings is higher than a chosen threshold. In both cases, embeddings are computed from a streaming audio window. However, we support the enrollment phrase specification either as audio or as text. In the case of text enrollment, we use Text To Speech (TTS) models~\cite{zalan2022audiolm,saeki2024extending} to generate audio, which is then used to generate the embedding. Furthermore, if multiple enrollment embeddings are available, the centroid of the enrollment embeddings is used to compute the cosine distance.

There has been an increasing amount of research to optimize custom KWS by building high quality speech embeddings. The most straightforward way is using an Automatic Speech Recognition (ASR) encoder~\cite{prabhavalkar2024extreme} to output a vector representation for an audio input window. The model usually has good accuracy for speech recognition and keyword matching, however the model size is usually too large to run on-device in a continuous manner.

In recent years, researchers have used transfer learning to extract speech embeddings from the hidden layers of models that are trained using classification tasks. In Lin et al.~\cite{lin2020training}, a speech encoder with five convolution layers is connected to different and independent decoders that perform classification tasks. The encoder parameters are shared when training different decoders and the output of the shared encoder is used as speech embedding. 
Similarly, Berg et al.~\cite{berg2021keyword} explores self-attention based encoding and Ding et al.~\cite{ding2022letr} built an encoder using a combination of convolution layers and a transformer architecture to perform classification tasks. Rybakov et al.~\cite{rybakov2020streaming} benchmarks the most popular architectures such as DNN, SVDF, LSTM, CRNN, MHAtt, etc. in the same setting. Although their models are device friendly, they are evaluated based on classification accuracy in the speech command dataset~\cite{warden2018speech}, which does not directly measure the quality of matching two separate utterances (i.e. test utterance and enrollment utterance). To evaluating keyword spotting quality in a setting closer to how users experience in the real world, we measure utterance matching accuracy directly by dividing the speech command dataset into enrollment dataset and testing dataset. In our experiments, the above classification model~\cite{lin2020training} does not perform well.

Triplet loss gained popularity for efficiently solving face identification problems, e.g. FaceNet~\cite{schroff2015facenet}. Since then, there has been more research on triplet loss in the speech domain. 
Bredin~\cite{bredin2017tristounet} and Song et al.~\cite{song2018triplet} use triplet-loss based learning approaches for the speaker diarization. Zhang and Koshida~\cite{zhang2017end} use triplet-loss for training speaker verification models. 
Recently, triplet loss has been applied to keyword spotting tasks. Sacchi et al.~\cite{sacchi2019open} and Chidhambararajan et al.~\cite{chidhambararajan2022efficientword} applied triplet loss on Open-Vocabulary Keyword Spotting, and built embeddings for texts and audio. However their sampling method is not efficient---each batch only samples one positive and one negative pair with the anchor utterance. This slows down the training and increases convergence instability because the selection of a single anchor utterance introduces high variances. Furthermore, their experiments were based on only a two layer GRU architecture. We experiment with different sizes of LSTMs~\cite{lstmref} and also apply state-of-the-art speech models including Conformer~\cite{gulati2020conformer}.

The main contributions of this paper include:
\begin{itemize}[leftmargin=*]
\item We apply \textit{generalized end-to-end loss} from the domain of speaker verification~\cite{wan2018generalized} to custom KWS tasks, simulating the two-stage process of runtime enrollment and verification during the training. This allows the training process to simulate the end user journey in the real world application. Unlike triplet loss, which compares a single anchor utterance against one positive utterance and one negative utterance, we build one enrollment centroid per phrase(a.k.a. keyword) in the batch, and compare it with all positive and negative utterances. This allows us to optimize matrix operations when computing the loss for a training batch for more efficient training. Moreover, since the enrollment centroids are constructed from multiple utterances containing the same phrase, the reduced sampling variances improve convergence stability. Therefore we noticed a fast and stable convergence in our experiments.

\item
Based on our literature review, to the best of our knowledge, we do not find a systematic and reliable way to evaluate custom KWS model performances. The triplet loss paper~\cite{sacchi2019open} uses fairly old WSJ dataset~\cite{paul1992design} collected in 1992, 
and Chidhambararajan's one-shot KWS~\cite{chidhambararajan2022efficientword} reports model results based on single threshold metrics from its TTS synthesized dataset and a small collected dataset with only 4 keywords. We design an end-to-end evaluation process that simulates the real world audio enrollment and custom keyword detections based on the commonly used speech command dataset~\cite{warden2018speech} in clean and noisy conditions. We define a set of metrics to directly measure keyword matching quality. 
We detail the evaluation process in the paper so it can be used in other KWS research.

\item 
We apply and tune a Conformer~\cite{gulati2020conformer} model for custom KWS, which beats the 7.5GB ASR encoder's AUC by 23.6\% using a 419KB quantized model. Our quantization method in Sec.~\ref{ssec:quantization} allows a low memory footprint model efficiently running on device in a continuous manner, while preserving the high accuracy. Our models are natively streamable. They are trained and exported using framework mentioned in Sec.~\ref{ssec:trainresource} with automatic streaming conversion features. Finally, we train and evaluate various sizes of LSTM and Conformer models, and show the tradeoff between model quality and model size. This allows the flexibility to choose the most suitable model given device type and runtime constraints.
\end{itemize}

The rest of the paper is organized as follows: We discuss related work and use them for our benchmark experiments in Section~\ref{sec:related_work}. \textit{generalized end-to-end loss} is applied to our different model architectures in Section~\ref{sec:method}. We design a new offline evaluation process simulating real world KWS environments in Section~\ref{sec:exps}, and discuss experiment's results in Section~\ref{sec:results}. The paper concludes in Section~\ref{sec:conclusion}.

\section{Model Baselines}
\label{sec:related_work}
\subsection{ASR Encoder for Keyword Spotting} 
ASR systems are designed to transcribe speech into text. In the context of keyword spotting, researchers have applied ASR encoders to build utterance embeddings. In Prabhavalkar et al.~\cite{prabhavalkar2024extreme}, the context length of ASR encoder can be up to 2.56 seconds and thus capture enough of the audio sequence to build a representation without needing temporal pooling. Given that ASR encoders are trained with a large amount of data and large models, it works very well on keyword spotting tasks. We use a pre-trained 7.5GB ASR encoder from this paper to benchmark our models.

\subsection{Speech Classification Encoder}
Researchers have been using transfer learning to extract speech embeddings from hidden layers of models trained for other tasks such as classifications. In Lin et al.~\cite{lin2020training}, a five layer convolution encoder is built and trained with 125 independent classifiers, sharing the encoder parameters during training. They use 200M short utterances (2s) mined from YouTube to train a 40-class model for each classifier. The output layer of the shared encoder is extracted as the speech embedding, published in site~\cite{kagglespeechemb}. We use a pre-trained 1.4 MB speech classification encoder in our experiment baseline.

\begin{figure*}[ht]
	\centering
	\includegraphics[width=0.52\textwidth]{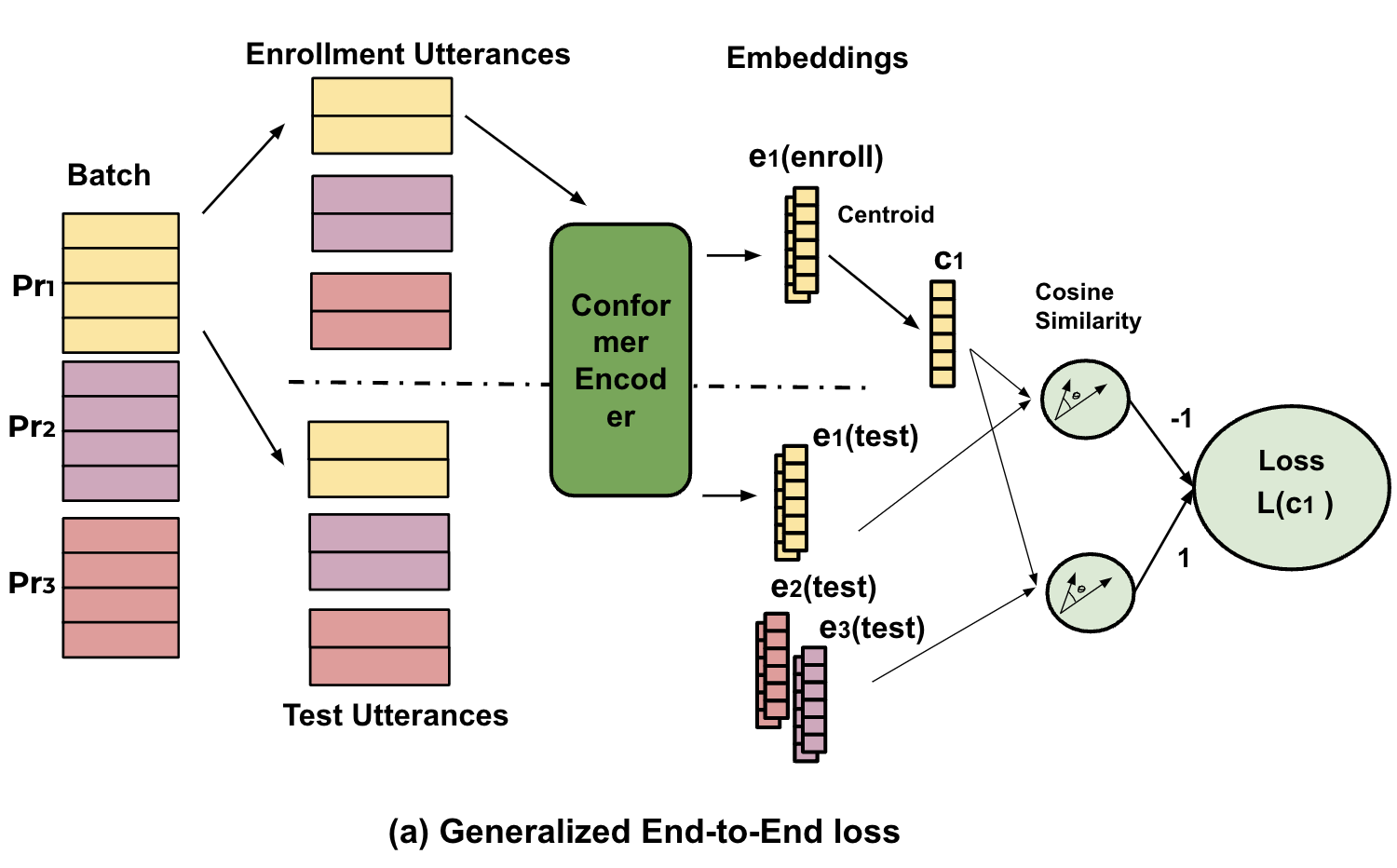}
	\includegraphics[width=0.45\textwidth]{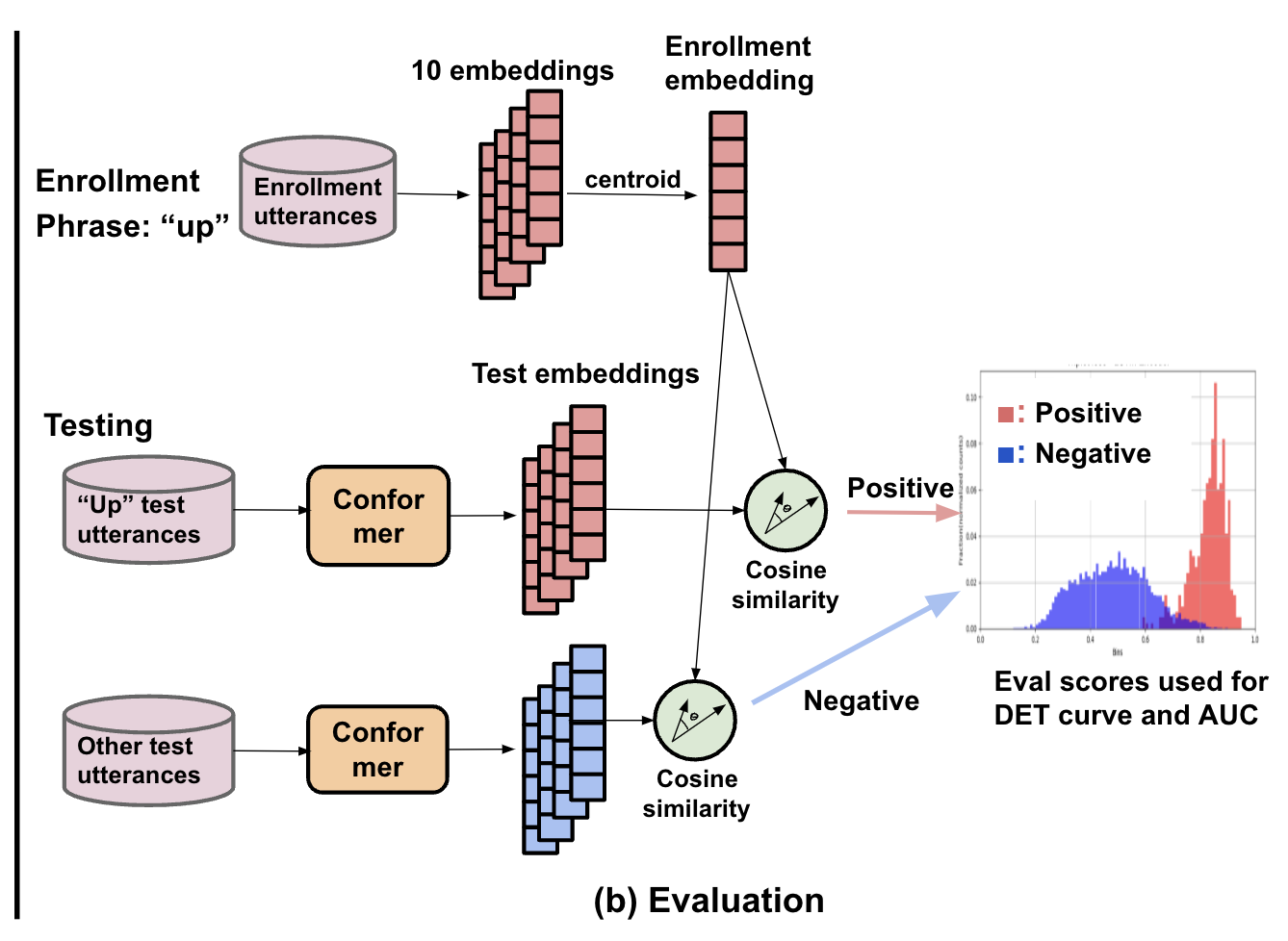}
	\caption{(a) An illustration of GE2E-KWS tasks in Section.~\ref{ssec:ge2e}. This example calculates the GE2E loss $\mathbf{L(c}_{1})$ for Phrase $\mathbf{Pr}_{1}$, in the context of phrase number $X=3$ and utterances per phrase $Y=4$ (splitted into 2 enrollment utterances and 2 testing utterances). Enrollment embeddings $\mathbf{e}_{1}\mathbf{(enroll)}$ and test embeddings $\mathbf{e}_{i}\mathbf{(test)}, i=1,2,3$ are extracted after conformer encoder. Then enrollment centroid $\mathbf{c}_{1}$ is computed and used to compare with different phrases' testing embeddings $\mathbf{e}_{i}\mathbf{(test)}, i=1,2,3$ by cosine similarity. Lastly GE2E loss $\mathbf{L(c}_{1})$ is computed by combining above similarities using Eq(\ref{eq:loss_fn}). (b) An example of the new evaluation process that simulates the real world serving environment.}
	\label{fig:abstract_graph}
\end{figure*}

\subsection{Triplet Loss Conformer}
\label{ssec:triplet_loss_baseline}
Triplet loss has been applied to optimize embedding representations for custom KWS from Sacchi et al.~\cite{sacchi2019open}. They construct the triplet loss by sampling an anchor utterance, and a positive utterance that has the same keyword as the anchor utterance, and a negative utterance otherwise. We use the their triplet loss function, and train the models with the same data and model architecture as the our treatment experiments to ensure fair comparisons.

\section{Proposed method}
\label{sec:method}

\subsection{Generalized End-to-End (GE2E) Loss} 
\label{ssec:ge2e}
We apply the \textit{generalized end-to-end} (GE2E) solution from the domain of speaker verification~\cite{wan2018generalized}, which splits all utterances in a batch into enrollment and test utterances. For each phrase, an enrollment centroid is built from all its enrollment utterances. This reduces training variance by building enrollment centroid from multiple utterances and by comparing it with all test utterance embeddings from the batch.

Fig.~\ref{fig:abstract_graph}(a) gives an example of the process calculating GE2E loss. Formally, In a batch we have $X$ different phrases $\mathbf{Pr}_{i}, i=1,2,..,X$, each phrase has $Y$ ($Y$ is an even number) utterances from different speakers and speaking styles. Let $\mathbf{e}_{ij}$ be the embedding for the $i$th phrase and the $j$th utterance. For each phrase $\mathbf{Pr}_{i}$, we compute an embedding centroid $\mathbf{c}_i$ using $Y/2$ utterances as enrollment while the remainder of the utterances for phrase $\mathbf{Pr}_{i}$ are later used as test utterances. This yields $X$ centroids, one for each phrase, with each centroid $\mathbf{c}_i$ being defined as follows:
\begin{align}
\mathbf{c}_i = \frac{1}{Y/2} \sum_{\substack{j=1 \\ j \;(\mathrm{mod}\; 2) \neq 0}}^{Y}{\mathbf{e}_{ij}}
\end{align}

Each centroid $\mathbf{c}_i$ is paired with positive example embeddings ($\mathbf{p_i}$)  which consist of the held-out test utterances that include phrase $\mathbf{Pr}_{i}$, and negative example embeddings ($\mathbf{n_i}$) that consist of all held-out test utterances from other phrases. These can be formally defined as follows:

\begin{align}
\mathbf{p}_{i} &= \{\mathbf{e}_{ij} \mid j \;(\mathrm{mod}\; 2) \neq 1 \}
\end{align}
\begin{align}
\mathbf{n}_{i} &= \{\mathbf{e}_{kj} \mid j \;(\mathrm{mod}\; 2) \neq 1, k=1,2,..,X \text{ and } k \neq i \}
\end{align}

We compute the GE2E loss using cosine similarity scores such that we (a) maximize score between a centroid and its positive examples, and (b) minimize score between a centroid and its negative examples. This can be formalized as follows for a centroid $\textbf{c}_i$: 

\begin{align}
\label{eq:loss_fn}
L(\textbf{c}_i) = \log{\sum_{n \in \textbf{n}_i}{\exp{\cos(\textbf{c}_i, n)}}} - \log{\sum_{p \in \textbf{p}_i}{\exp{\cos(\textbf{c}_i, p)}}}
\end{align}

GE2E has two advantages:
\begin{itemize}[leftmargin=*]
\item Increases convergence stability by reducing the sampling variance and building enrollment centroids from multiple utterances.
\item Increases computational efficiency by formatting embedding similarity computations of \{enrollment, test utterance\} pairs into matrix operations.
\end{itemize}

\subsection{Model architectures} 

\subsubsection{LSTM Model}
LSTM~\cite{lstmref} is a commonly used RNN to model sequential features like audio sequences. We used the standard three-layer LSTM architecture with different hidden layer and output layer dimensions to optimize the results and tradeoff model performance and model size.

\subsubsection{Conformer Model}
The Conformer model was proposed by Gulati et al.~\cite{gulati2020conformer} to tackle ASR problems and has been the SOTA for ASR encoders. It combines the strengths of Transformers, which capture the global interactions of audio sequences, and strengths of Convolution Neural Networks (CNNs), which exploit local features effectively. In this paper, we used the standard model architectures from the above paper. We have tuned the number of Conformer blocks, the number of heads in Multi Head Self Attention (MHSA) and the output embedding sizes etc. to trade off between model performance and model sizes.

\subsection{Quantization} 
\label{ssec:quantization}
We use the dynamic range quantization from TensorFlow Lite~\cite{quantizationref}. This dynamically quantizes weights and activations to 8-bits of precision based on the ranges of those weights and activations.

\section{Experiment setup and Evaluation}
\label{sec:exps}

\subsection{Training Dataset and Infra}
\label{ssec:trainresource}

Our training data are processed from the MSWC open source dataset~\cite{mazumder2021multilingual}, which includes 38k different phrases and 5.3M training utterances from approximately 27k speakers. The audio sequences are processed into 40-dimensional spectral features with each frame length of 25ms. In a batch we randomly choose 8 phrases, and sample 10 utterances per phrase. For each phrase, we compute the GE2E loss detailed in Fig.~\ref{fig:abstract_graph}(a) and Sec.~\ref{ssec:ge2e}. For training framework, our model developments are based on Tensorflow/Lingvo~\cite{Shen2019LingvoAM} with advantages of automatic streaming inference conversion.


\subsection{Evaluation dataset and process}
We use the test split of Speech command dataset~\cite{warden2018speech}, which has 11k utterances across 35 phrases from different speakers. Shown in Fig.~\ref{fig:abstract_graph}(b), for a given phrase (e.g. ``up"), we randomly select 10 utterances as an enrollment dataset and use the remaining utterances for testing purposes. The enrollment centroid is aggregated from 10 enrollment utterance embeddings and is used to compare with all testing utterance embeddings by cosine similarity. In this process, we aim to achieve the same goals as in real world serving environments: The embedding similarity should be high if the enrollment centroid and test utterances are from the same phrase, and low otherwise. In addition, we apply 3-way MTR of 3db to 15db background noise to the utterances. Therefore our experiments are conducted in both clean and noisy conditions.

\subsection{Evaluation metrics}
\label{ssec:metrics}


\begin{figure}[!ht]
	\centering
	\includegraphics[width=4.25cm, height=4.25cm]{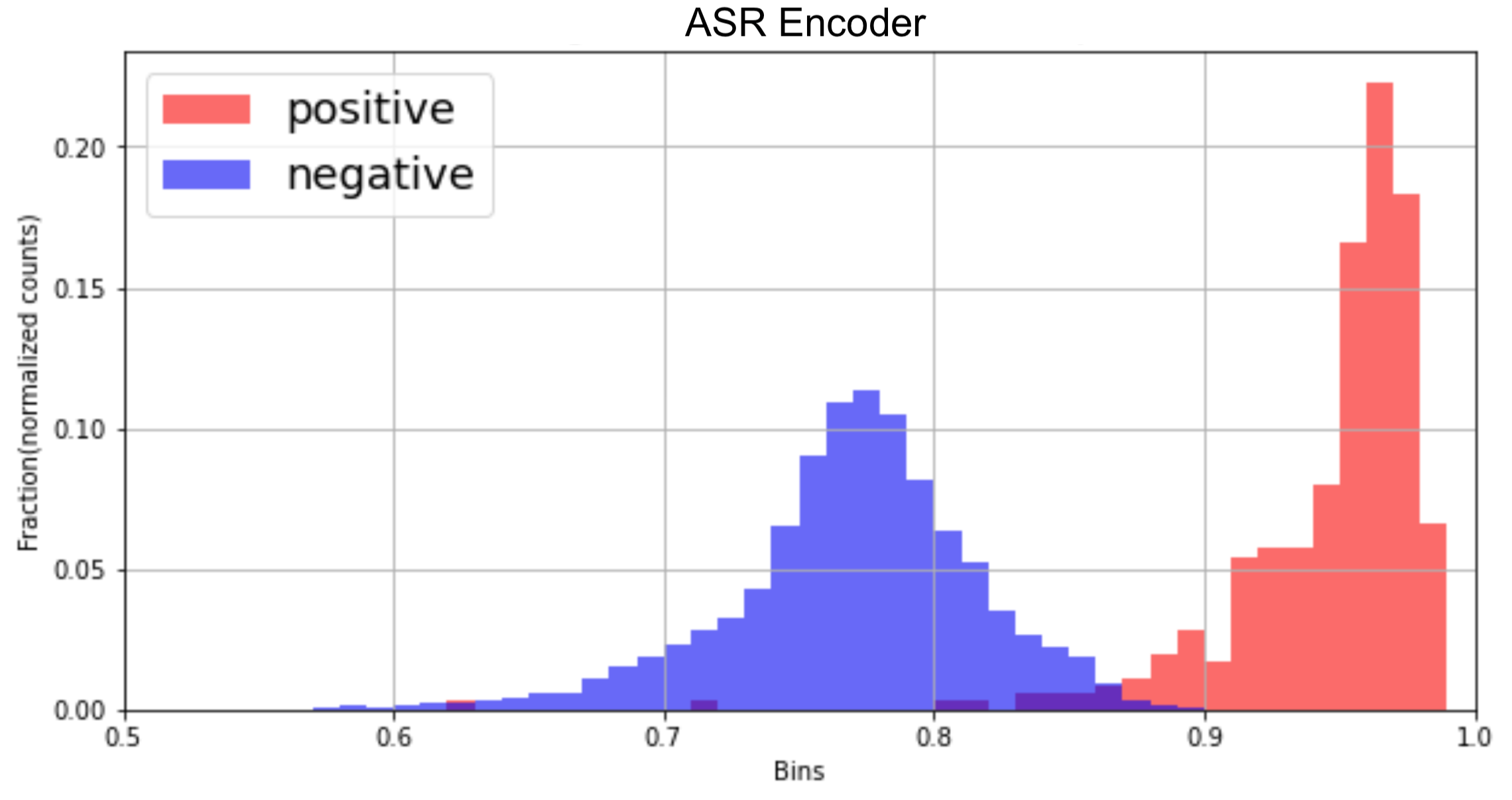}
	\includegraphics[width=4.2cm, height=4.2cm]{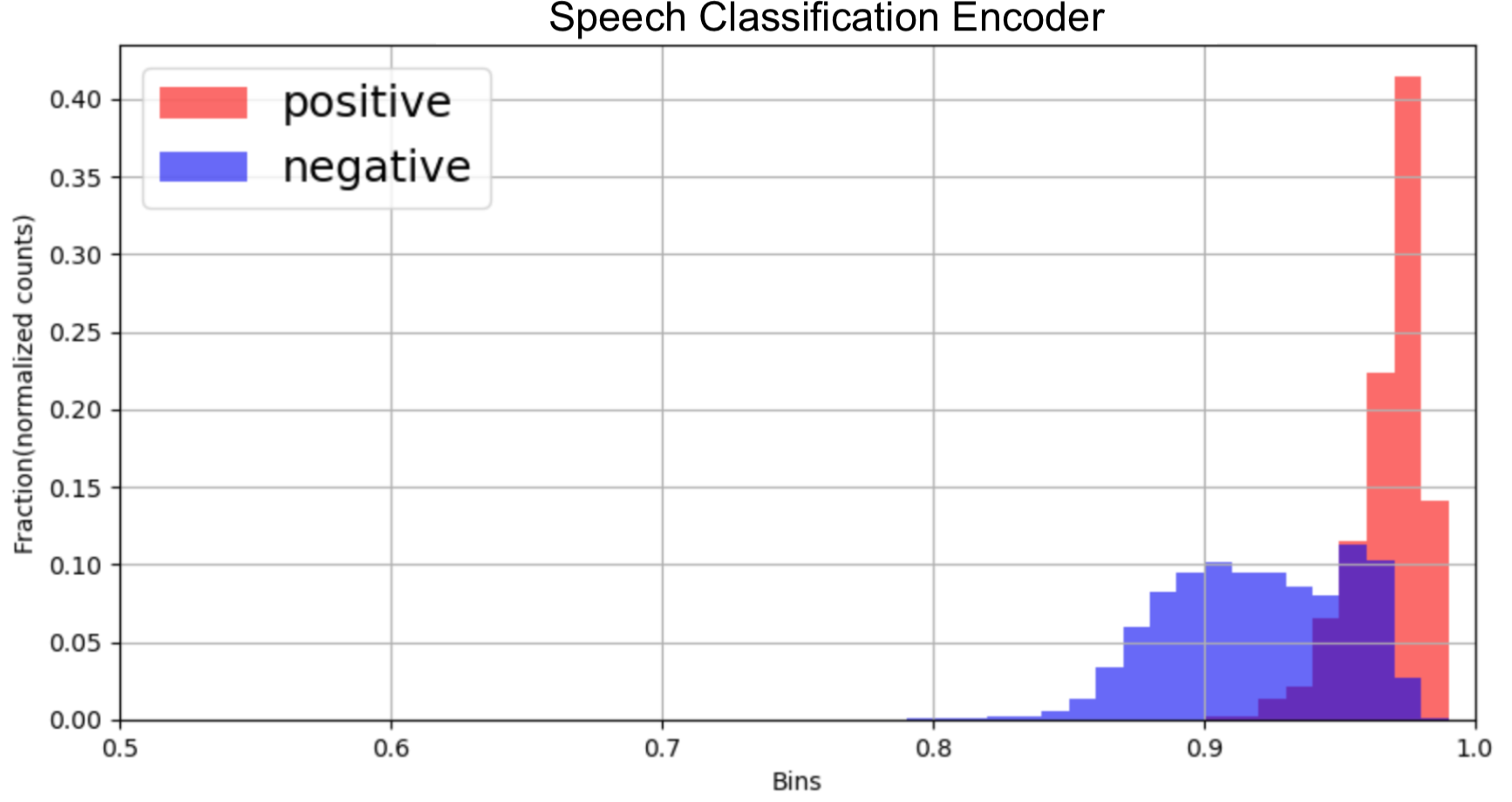}
	\includegraphics[width=4.25cm, height=4.25cm]{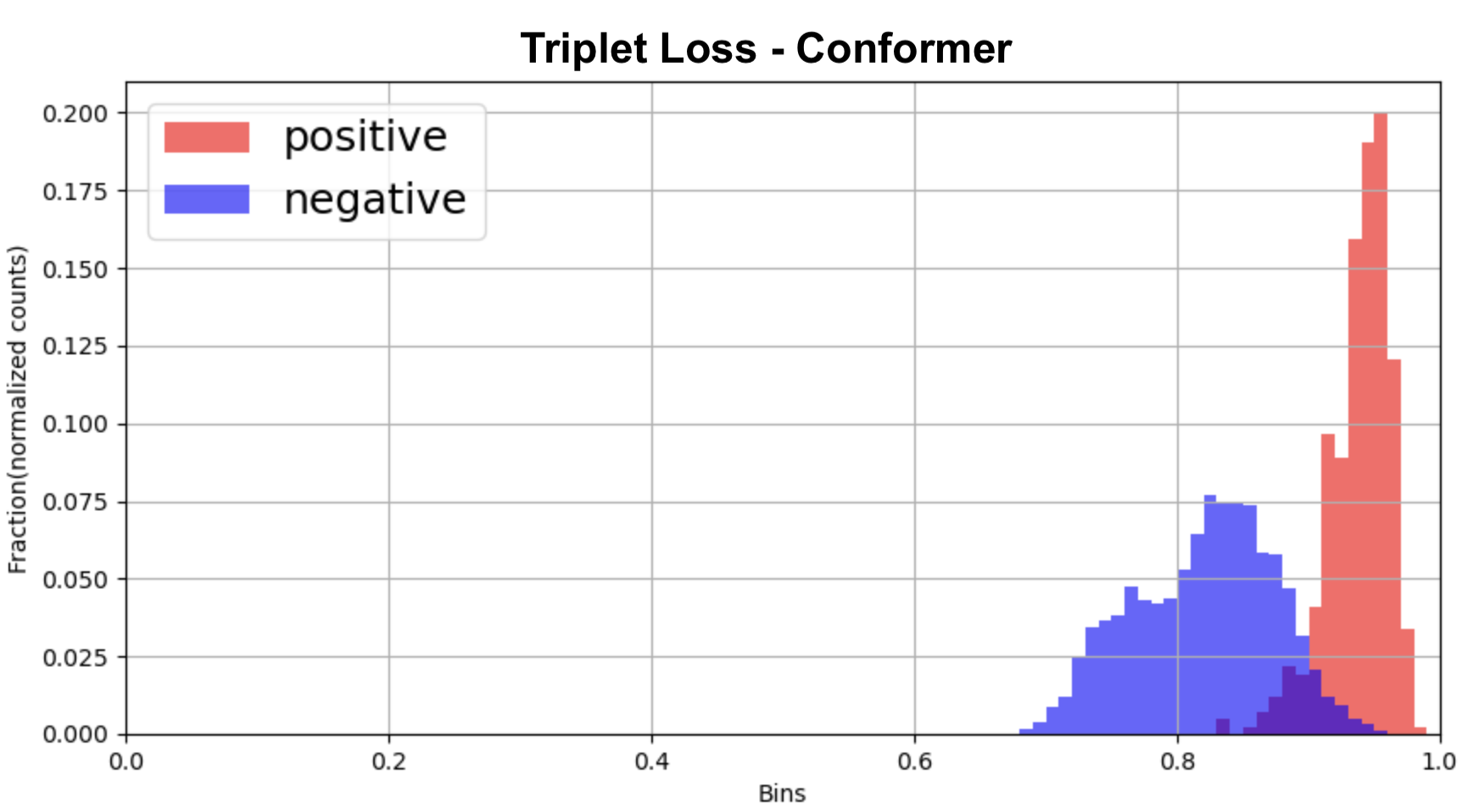}
	\includegraphics[width=4.25cm, height=4.25cm]{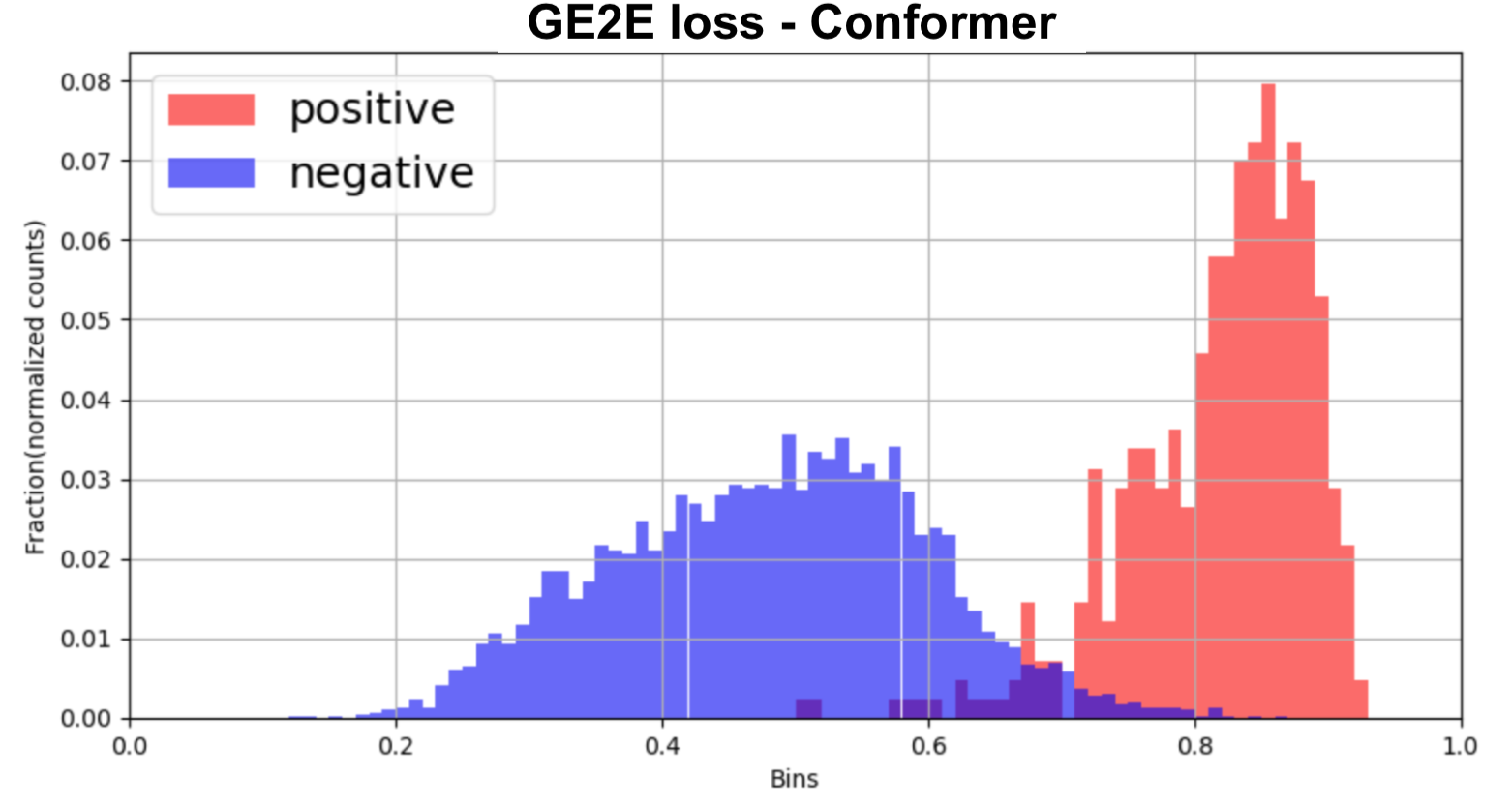}
	\caption{Similarity score histograms for different models. The enrollment-to-test utterance similarity scores are grouped into 100 bins from 0 to 1 with a stride of 0.01. The counts are normalized to probability distributions. The phrase used here is ``up'', a word not present in the training data.}
	\label{fig:hists}
\end{figure}

\begin{figure}[!ht]
	\centering
	\includegraphics[width=4.25cm, height=4.25cm]{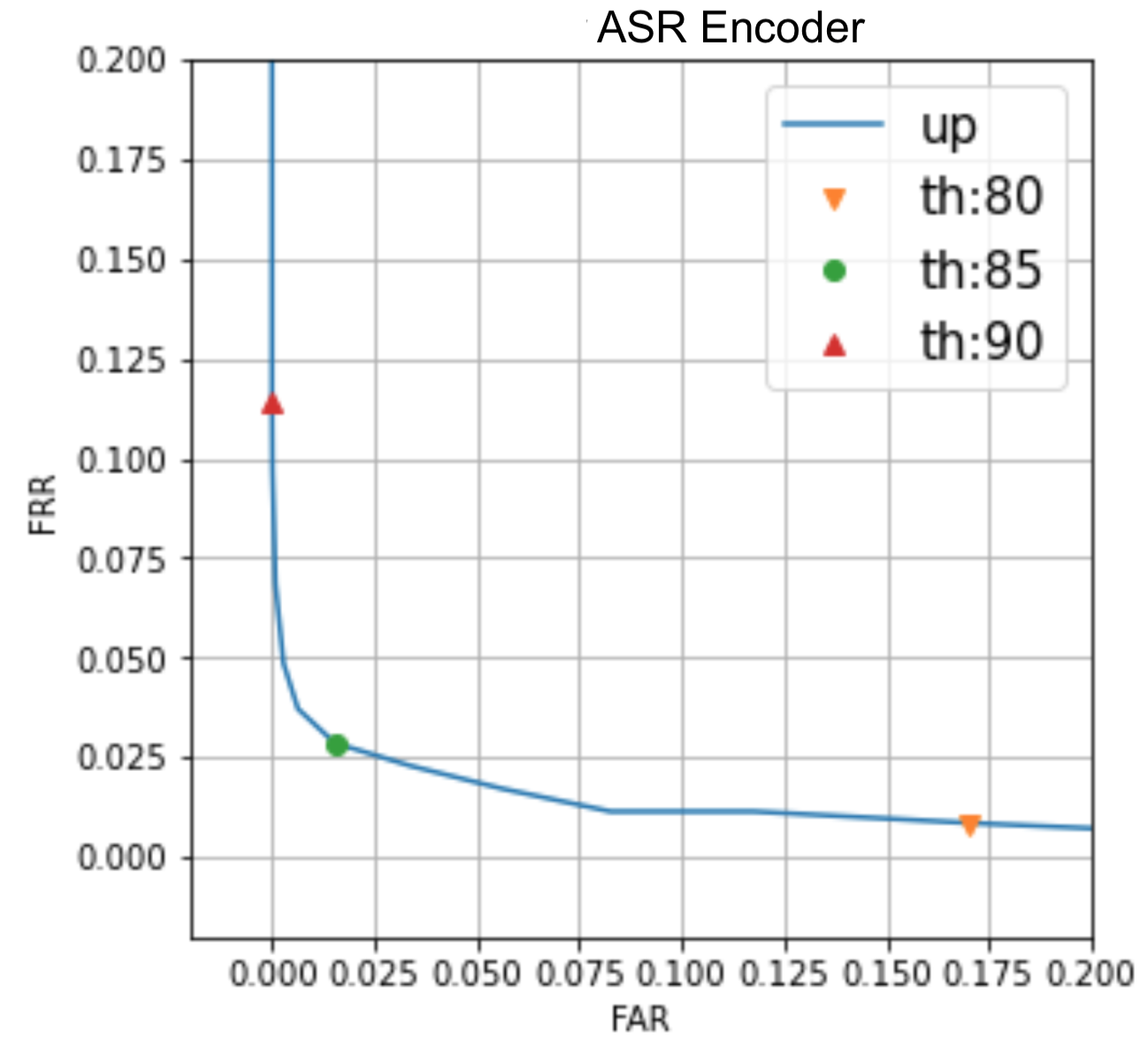}
	\includegraphics[width=4.25cm, height=4.25cm]{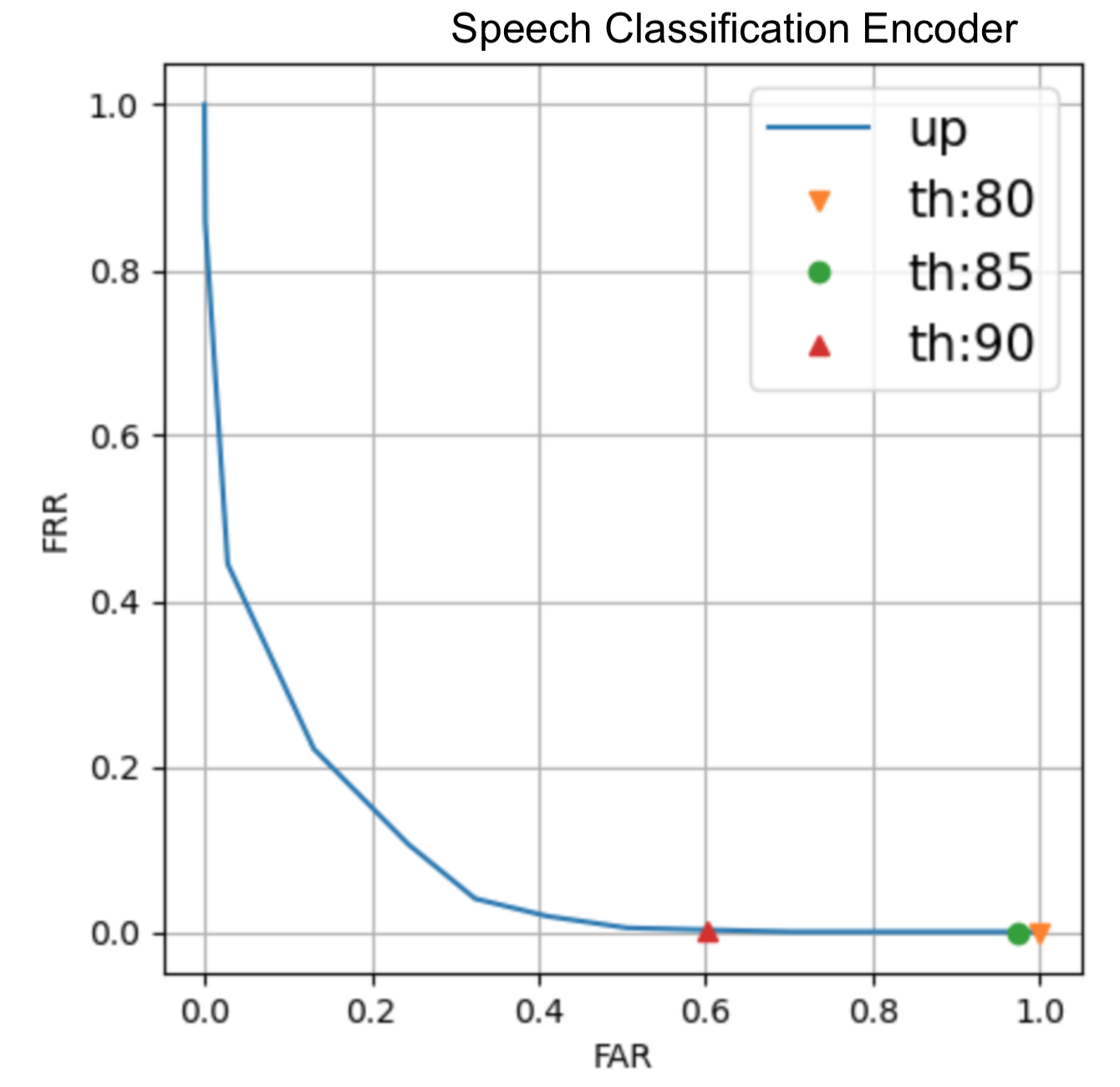}
	\includegraphics[width=4.25cm, height=4.25cm]{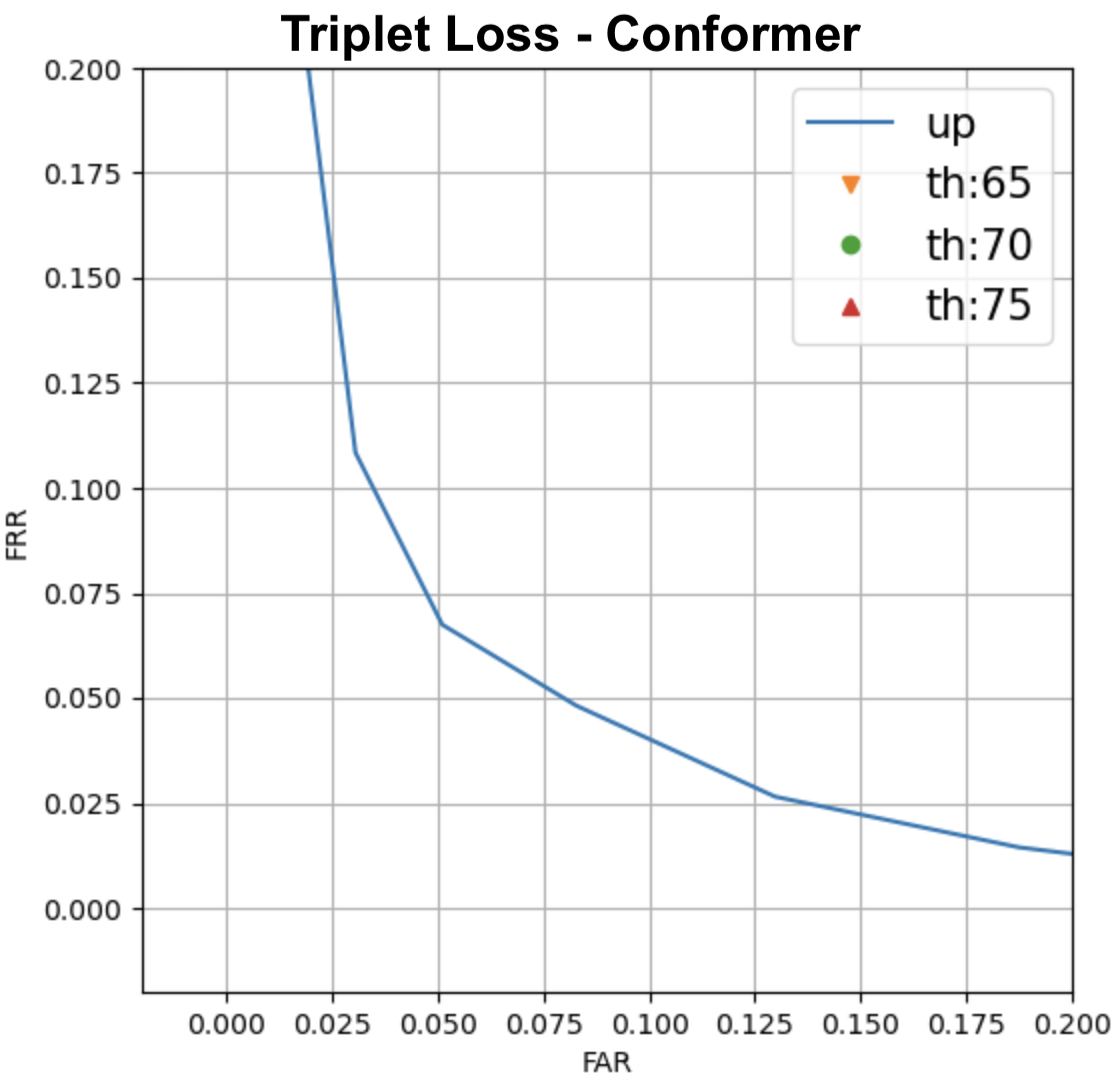}
	\includegraphics[width=4.25cm, height=4.25cm]{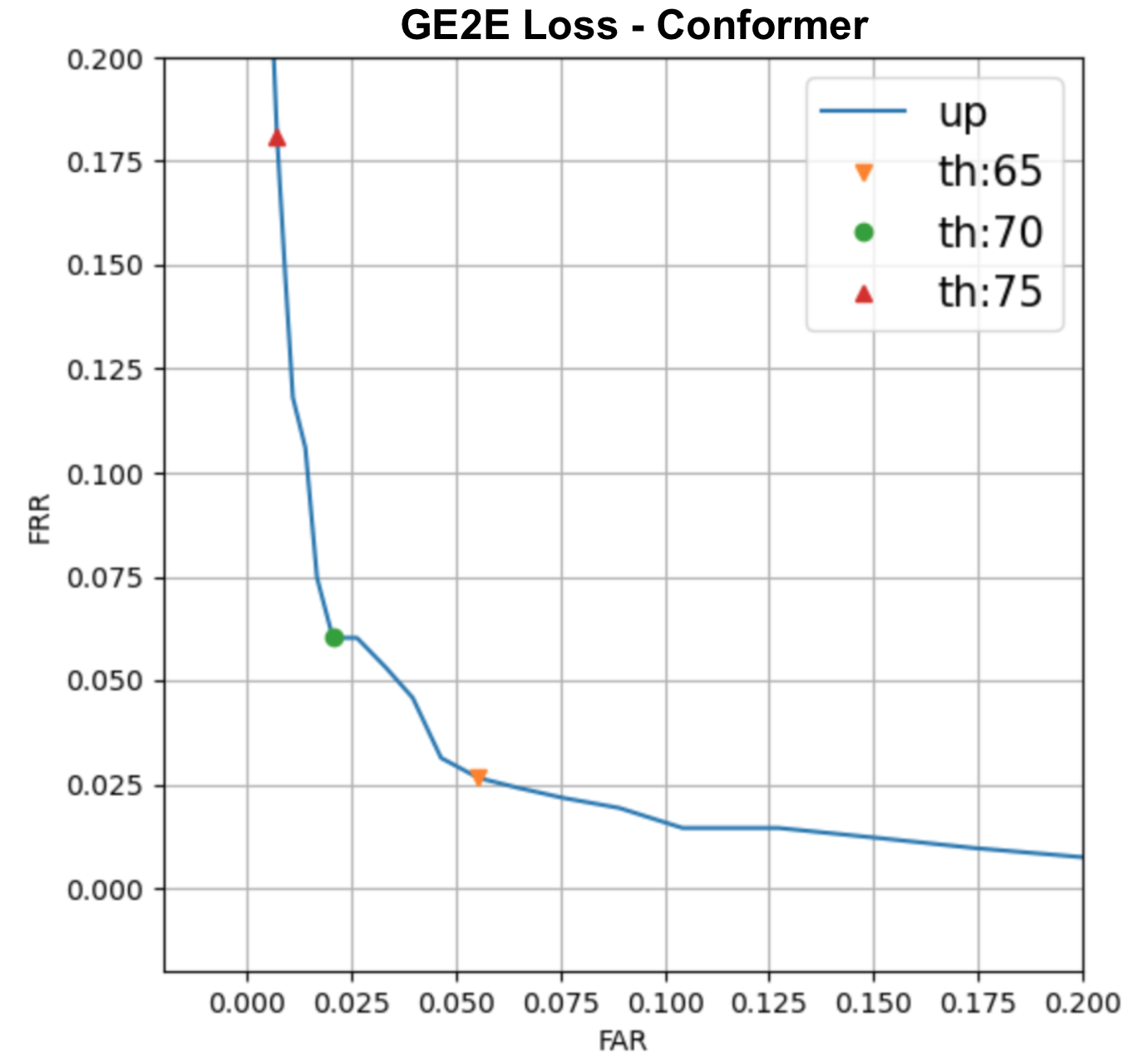}
	\caption{DET(FRR-FAR) curves for different models. Example thresholds are marked in the curves to show the corresponding FRRs and FARs. The axis scales are adjusted to accommodate different model results. The phrase used here is ``up'', a word not present in the training data.}
	\label{fig:rocs}
	\vspace{-3mm}
\end{figure}

As mentioned above we have 35 enrollment centroids corresponding to 35 phrases. Illustrated in Fig.~\ref{fig:abstract_graph}(b), for each phrase, we compare its enrollment centroid against all testing utterance embeddings, and plot their cosine similarity scores into the histogram. Fig.~\ref{fig:hists} shows the histogram of different models for phrase ``up''.  In each histogram, if a test utterance comes from the same phrase of enrollment utterances, we treat it as a true positive example and count its similarity score into red frequency bars. If the test utterance is from a different phrase, it is treated as true negative and the similarity score is counted into blue frequency bars. We normalize the frequency bars into probability distributions as there are many more negative pairs.

A perfect model will have a threshold that can perfectly separate the scores from true positive examples and true negative examples. However in the overlapping areas, the model struggles to make the right decision. When applying different thresholds in the score histogram, from 0 to 1 with a stride of 0.01, we calculate FAR and FRR for each threshold and plot the Detection Error Tradeoff (DET) curves. Fig.~\ref{fig:rocs} shows the DET curves for different models for the phrase ``up''. 

To directly compare model performance using numeric metrics, we use the Area Under the DET Curve (AUC) and Equal Error Rate (EER) of the DET curves. In this context, EER is the value of FAR and FRR when they are equivalent. In both cases, lower AUC and lower EER indicate better model performance. Since each phrase has its own DET curve, the AUC and EER measure the model performance when matching a particular phrase. In order to have metrics that holistically evaluate a model independent of phrase, we compute aggregated AUC and EER by averaging metrics over different phrases. For the rest of the paper, unless otherwise noted, metrics such as AUC, EER and DET refer the aggregated version. In addition, AUC is a numeric value from 0 to 1. However for easier readings, AUCs are formatted with ``\%" in most places of the paper.

\section{Results}
\label{sec:results}

\subsection{Results for Different Models}
\label{ssec:results_baseline_models}
As mentioned in Sec.~\ref{sec:related_work}, we compare our GE2E conformer model with a pre-trained 7.5GB ASR encoder, a pre-trained 1.4MB speech classification encoder and a triplet loss based conformer model sharing the same model architecture, training data and similar training epochs. Our raw conformer model has size 2.8MB and is quantized into 419KB using techniques in Section.\ref{ssec:quantization}. The evaluations are based on quantized models.

\begin{figure}[!ht]
	\centering
	\includegraphics[width=0.5\textwidth]{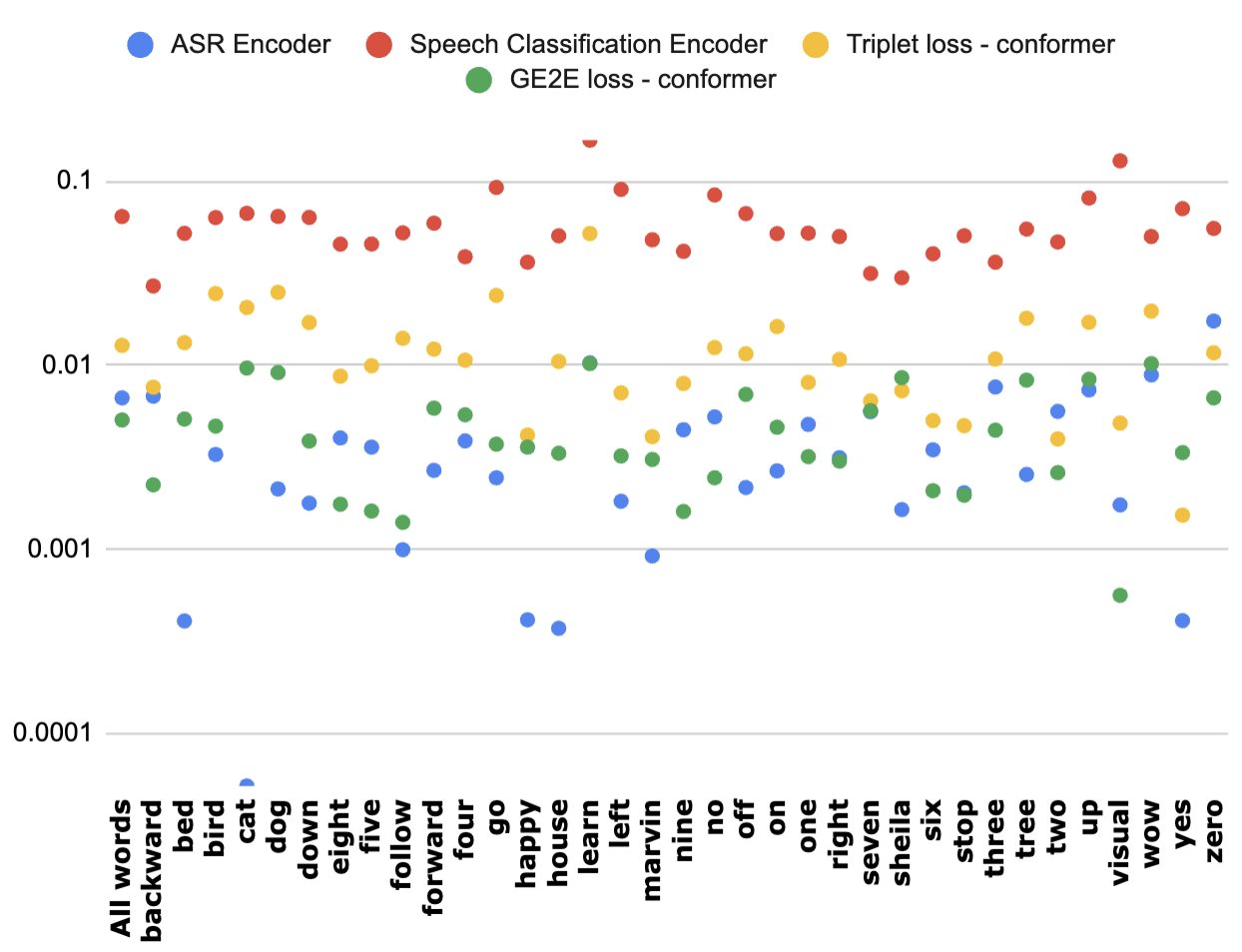}
	\caption{Word level AUC and aggregated AUC comparisons across different models. ``no'', ``go'', ``on'', and ``up'' phrases do not appear in the training data.}
	\label{fig:scatters}
\end{figure}

The above Fig.~\ref{fig:hists} and Fig.~\ref{fig:rocs} show the score histograms and Detection Error Tradeoff (DET) curves for the phrase ``up'' across different models. Table.\ref{tab:baseline_comps} shows AUCs for few other individual phrases and also an all-words aggreated AUC defined in Section.\ref{ssec:metrics}. To better visualize model performances, AUC values are plotted in Fig.~\ref{fig:scatters} for all 35 words in the speech command dataset. The leftmost datapoints measure the overall model performance from ``All words AUC''. From the above table and plots, we can see the ASR model performs well with an AUC of 0.66\%, but with a big size 7.5GB. The speech classification encoder has a device-friendly model size but it performs poorly with an AUC of 6.44\%. The triplet loss conformer baseline with raw size 2.8MB and quantized model 419KB has 1.283\% AUC. Noticeably, our GE2E loss model that has the same size as the triplet loss conformer performs the best here with an AUC of 0.504\%, relatively 23.6\% better than ASR encoder and 60.7\% better than triplet loss conformer.

\begin{figure}[!ht]
	\centering
	\includegraphics[width=0.37\textwidth]{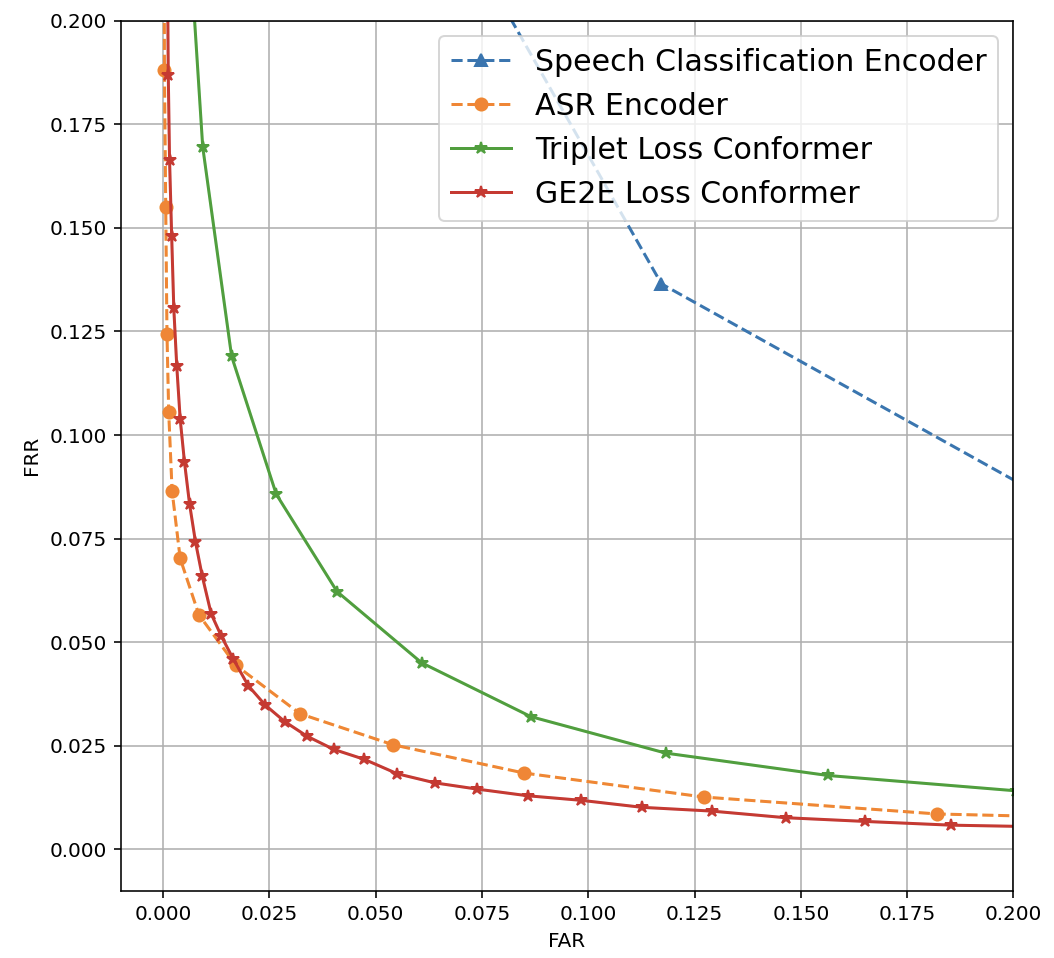}
	\caption{Aggregated DET curves for different models}
	\label{fig:roc_4models}
\end{figure}

In Fig.~\ref{fig:roc_4models}, the DET curves for those models are plotted to show the accuracy at different operating points. The DET curves are aggregated over all evaluation phrases by averaging their FRRs and FARs.

\begin{table}[!htbp]
\centering
\resizebox{9cm}{!}{
\begin{tabular}{l|l|l|l|l}
Keyword & ASR Encoder & \begin{tabular}[c]{@{}l@{}}Speech \\Classification\\ Encoder\end{tabular} & \begin{tabular}[c]{@{}l@{}}Regular \\ Triplet loss \\Conformer\end{tabular} & \begin{tabular}[c]{@{}l@{}}GE2E loss\\ Conformer \end{tabular} \\ \cline{1-5} 
Model size  & 7.5GB      & 1.4MB                        & 419KB               & 419KB   \\ \cline{1-5} 
Training method  & Pre-trained      & Pre-trained                        & 15 epochs               & 13 epochs   \\
\cline{1-5}
\textbf{AUC:All words} & \textbf{0.66}\%      & \textbf{6.44}\%                        & \textbf{1.283}\%               & \textbf{0.504}\%   \\ \cline{1-5}
AUC:backward  & 0.68\%      & 2.70\%                        & 0.76\%               & 0.22\%   \\
AUC:bed       & 0.04\%      & 5.21\%                        & 1.33\%               & 0.51\%   \\
AUC:bird      & 0.33\%      & 6.35\%                        & 2.45\%               & 0.46\%   \\
AUC:no       & 0.52\%      & 8.43\%                        & 1.25\%               & 0.24\%   \\
AUC:up        & 0.73\%      & 8.11\%                        & 1.71\%               & 0.83\%  
\end{tabular}}
\caption{Word level AUC and aggregated AUC comparisons for baseline models in Sec.\ref{sec:related_work} and the GE2E loss conformer model. The full list of words is truncated due to space limit. For Conformer models, the evaluations are based on the quantized version(419KB) of original conformer models (2.8MB).}
\label{tab:baseline_comps}
\end{table}

\subsection{Results for Different LSTM and Conformer Model Sizes}
\begin{figure}[!ht]
	\centering
	\includegraphics[width=0.37\textwidth]{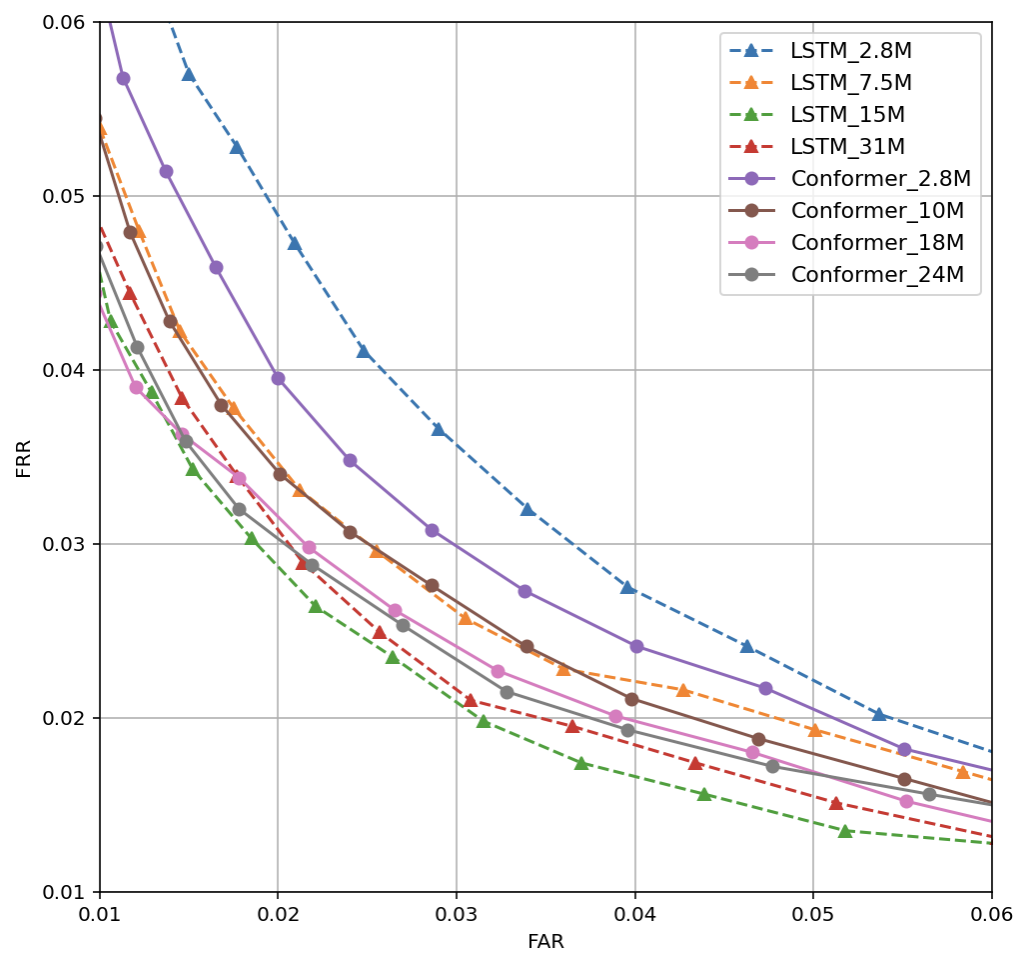}
	\caption{DET plots for various Conformer and LSTM Models.}
	\label{fig:various_size_rocs}
\end{figure}

Given GE2E loss outperforms triplet loss model and other baseline models, we tune GE2E loss based LSTM and Conformer model architectures to understand how model performance varies with model size. For the LSTM model, we retain three stacking layers to capture the sequence dependencies, and increase the hidden layer and output layer dimensions. As shown in top four rows in Table~\ref{tab:various_size_models} and triangle decorated dotted curves in Fig.~\ref{fig:various_size_rocs}, the model performance achieves optimal result at 15MB size (1.2MB quantized) with our training data recipe. However the smaller model might still be preferred if a lower memory footprint or latency is required in the production environment.

We did further parameter tuning for the conformer models as well, including experimenting with 5--12 conformer blocks, various number of heads in MHSAs, and various positional encoding and attention output layer dimensions. As shown in bottom four rows in Table.~\ref{tab:various_size_models} and solid curves in Fig.~\ref{fig:various_size_rocs}, the model performance approximates the optimal and plateaus when increasing the size to 18MB (1.8MB with quantizations). Note the 24MB model has slight better AUC but its DET curve is entangled with the 18MB model so we will not conclude who is better. Besides, the above results show that conformer has better result than LSTM at small model sizes(i.e. 2.8MB raw model size). As model size increases, the LSTM model performs slightly better in clean conditions, however the conformer models are better under noisy conditions because of the ability of detecting noise from local contexts.

\begin{table}[!htbp]
\centering
\resizebox{8cm}{!}{
\begin{tabular}{l|l|l|l|l}
\begin{tabular}[c]{@{}l@{}}Model size\\ (raw)\end{tabular} & \begin{tabular}[c]{@{}l@{}}Model size\\ (quantized)\end{tabular} & \begin{tabular}[c]{@{}l@{}} EER Clean\\ condition \end{tabular} & \begin{tabular}[c]{@{}l@{}} EER Noisy\\ condition\end{tabular} & AUC     \\ \cline{1-5}
2.8MB       & 250KB                   & 3.17\%              & 12.71\%               & 0.568\% \\ 
7.5MB       & 630KB                   & 2.62\%              & 10.65\%               & 0.444\% \\ 
15MB        & 1.2MB                  & 2.32\%              & 10.46\%              & 0.344\% \\ 
31MB        & 2.7MB                  & 2.45\%              & 10.91\%               & 0.381\% \\ 
\cline{1-5}
2.8MB       & 419KB                   & 2.94\%              & 9.69\%              & 0.504\% \\ 
10MB        & 1.2MB                   & 2.75\%                & 10.49\%               & 0.416\% \\ 
18MB        & 1.8MB                   & 2.5\%                 & 9.21\%                & 0.389\% \\ 
24MB        & 2.4MB                   & 2.51\%                & 8.98\%                & 0.382\%
\end{tabular}}
\caption{EER and AUC for different sizes of LSTM models (top four rows) and Conformer models (bottom four rows). Note the quantized model size also includes metadata and state buffers, which can vary upon model architectures.}
\label{tab:various_size_models}
\end{table}

\section{Conclusion}
\label{sec:conclusion}

This paper's main contributions include: (1) first time applying GE2E loss to simulate the two-stage process of runtime enrollment and verification during the KWS training and (2) design an end-to-end evaluation process that mimics the real world audio enrollment and custom keyword detection using the latest speech command dataset. On top of improved training speed and convergence stability, our GE2E loss and 419KB quantized conformer model beats the 7.5GB ASR encoder by 23.6\% on phrase averaged AUC and beats the same size triplet loss baseline by 60.7\%. Furthermore, our models are natively streamable and designed to run continuously on-device with no retraining needed for new keywords(zero-shot).

\clearpage
\bibliographystyle{IEEEbib}
\bibliography{strings,refs}

\end{document}